\documentclass[10pt]{article}

\usepackage{amssymb,amsmath,amsfonts,amsthm,a4}

\newtheorem{theorem}{Theorem}[section]
\newtheorem{lemma}[theorem]{Lemma}
\newtheorem{assumption}[theorem]{Assumption}

\theoremstyle{remark}
\newtheorem*{remark*}{Remark}
\newtheorem*{remarks*}{Remarks}

\newcommand{\RR}{\mathbb{R}}
\newcommand{\CC}{\mathbb{C}}

\newcommand{\dd}{\mathrm{d}}
\newcommand{\En}{\mathcal{E}}
\newcommand{\Nn}{\mathcal{N}}
\newcommand{\DD}{\Delta}
\newcommand{\Tm}{\sqrt{-\DD + m^2}  } 
\newcommand{\T}{\sqrt{-\DD}   } 
\newcommand{\Tpm}{\sqrt{p^2 + m^2}  }

\newcommand{\ie}{i.\,e.}
\newcommand{\eg}{e.\,g.}

\newcommand{\Hhalf}{H^{1/2}}
\newcommand{\eps}{\varepsilon}

\newcommand{\dt}{\partial_t}
\newcommand{\Lp}[1]{L^{#1}}
\newcommand{\Hs}[1]{H^{#1}}

\newcommand{\diff}{\, \mathrm{d} }
\newcommand{\inner}[2]{  \langle #1, #2  \rangle }
\newcommand{\innerb}[2]{ \big \langle #1, #2 \big \rangle }
\newcommand{\Ctest}{C^\infty_{c}}
\newcommand{\Vs}{V_{\mathrm{s}}}
\newcommand{\Nc}{N_{\mathrm{c}}}

\numberwithin{equation}{section}

\begin{document}

\title{{\bf Blow-Up for Nonlinear Wave Equations \\ describing Boson Stars}}
\author{J\"urg Fr\"ohlich\thanks{Institute for Theoretical Physics, ETH Z\"urich, Switzerland. E-Mail: {\tt juerg@itp.phys.ethz.ch}} \and Enno Lenzmann\thanks{Department of Mathematics, ETH Z\"urich, Switzerland. E-Mail: {\tt lenzmann@math.ethz.ch}}}

\maketitle

\begin{abstract}
We consider the nonlinear wave equation 
\[ i \partial_t u= \sqrt{-\Delta + m^2} u - (|x|^{-1} \ast |u|^2) u \quad \mbox{on $\RR^3$} \]
modelling the dynamics of (pseudo-relativistic) boson stars. For spherically symmetric initial data, $u_0(x) \in C^\infty_c(\RR^3)$, with negative energy, we prove blow-up of $u(t,x)$ in $H^{1/2}$-norm within a finite time. Physically, this phenomenon describes the onset of ``gravitational collapse'' of a boson star. We also study blow-up in external, spherically symmetric potentials and we consider more general  Hartree-type nonlinearities. As an application, we exhibit instability of ground state solitary waves at rest if $m=0$.  
\end{abstract}


\section{Introduction}

In this paper, we prove blow-up of solutions of the nonlinear wave equation 
\begin{equation} \label{eq-nls-intro}
 i \partial_t u = \Tm u - \big ( \frac{1}{|x|} \ast |u|^2 \big ) u \quad \mbox{on $\RR^3$}
\end{equation}
arising as an effective description of pseudo-relativistic {\em boson stars}, as recently shown in \cite{Elgart+Schlein2005, Schwarz2006}. Here $u(t,x)$ is a complex-valued wave field (a one-particle wave function). The operator $\sqrt{-\Delta+m^2}$, which is defined via its symbol $\sqrt{k^2 + m^2}$ in Fourier space, describes the kinetic and rest energy of a relativistic particle of mass $m \geq 0$, and $|x|^{-1}$ is the Newtonian gravitational potential in appropriate physical units. Moreover, the symbol $\ast$ stands for convolution on $\RR^3$.

Apart from its applications in theoretical astrophysics, equation (\ref{eq-nls-intro}) is of considerable interest from the PDE's point of view: It is a ``semi-relativistic'' nonlinear Schr\"odinger (or Hartree) equation with focusing, $L^2$-critical nonlinearity. In particular, there exist {\em travelling ground state solitary waves;} an extensive study can be found in \cite{FJL2005}. 

The purpose of the present paper is to address the problem of proving blow-up of solutions, and we will establish the following result: Any spherically symmetric initial datum, $u_0(x) \in \Ctest(\RR^3)$, with negative energy, 
\begin{equation}
\En(u_0) < 0,
\end{equation}
gives rise to a solution, $u(t,x)$, of (\ref{eq-nls-intro}) that blows up within a finite time, \ie, we have that
\begin{equation} \label{eq-blowup}
\lim_{t \nearrow T} \| u(t, \cdot) \|_{H^{1/2}} = \infty , \quad \mbox{for some $0 < T < \infty$}.
\end{equation}
Here the energy functional, $\En(u)$, is given by
\begin{equation}
\En(u) = \frac{1}{2} \int_{\RR^3} \overline{u} \Tm u \diff x - \frac{1}{4} \int_{\RR^3} \big ( \frac{1}{|x|} \ast |u|^2 \big ) |u|^2 \diff x,
\end{equation}
and $\| \cdot \|_{H^{1/2}}$ in (\ref{eq-blowup}) denotes the norm of the Sobolev space $\Hhalf(\RR^3)$. In more generality, this blow-up result is described in Theorem \ref{th-blowup} below, which also takes external potentials and other types of Hartree nonlinearities into account. In physical terms, finite-time blow-up of $u(t,x)$ is indicative of the onset of {\em ``gravitational collapse''} of a boson star modelled by (\ref{eq-nls-intro}). 

We begin with a brief recapitulation of some important results for (\ref{eq-nls-intro}) that have been derived so far. As shown in \cite{Lenzmann2005LWP}, the Cauchy problem for (\ref{eq-nls-intro}) is locally well-posed for initial data, $u_0(x)$, that belong to the Sobolev space $\Hs{s}(\RR^3)$, with $s \geq 1/2$, where $\Hhalf(\RR^3)$ is the energy space for (\ref{eq-nls-intro}). Moreover, we have that $u(t,x)$ extends to all times, $t \geq 0$, provided that the initial datum satisfies
\begin{equation} \label{eq-Nc}
\int_{\RR^3} |u_0|^2 \diff x <  \Nc,
\end{equation}
where $\Nc > 4/\pi$ is a universal constant; see \cite{Lenzmann2005LWP} for more details. In particular, we point out that condition (\ref{eq-Nc}) implies that $\En(u_0) > 0$ holds. Thus, blow-up phenomena for (\ref{eq-nls-intro}) can only occur for large initial data, $u_0(x)$, that do {\em not} satisfy condition (\ref{eq-Nc}). Concerning its physical interpretation, the universal constant $\Nc$ (which defines the scale of large and small initial data) can be viewed as the {\em ``Chandrasekhar limit mass''} for boson stars modelled by (\ref{eq-nls-intro}). Furthermore, the boson star equation (\ref{eq-nls-intro}) has been rigorously derived in \cite{Elgart+Schlein2005} from many-body quantum mechanics; we refer to \cite{Lieb+Yau1987} for an earlier result on the time-independent problem. 

As an application of our blow-up result, we can study the {\em stability of ground state solitary waves} (at rest),
\begin{equation}
u_{\mathrm{sol}}(t,x) = e^{i\omega t} Q(x),
\end{equation}
where $Q(x) \in \Hhalf(\RR^3)$, with $Q \not \equiv 0$, is nonnegative (up to a constant phase) and satisfies
\begin{equation}
\Tm  Q - \big ( \frac{1}{|x|} \ast |Q|^2 \big ) Q = -\omega Q,
\end{equation}
for some $\omega > 0$. Indeed, existence and spherical symmetry of solutions $Q(x)$, which we define as minimizers of $\En(u)$ subject to $\int_{\RR^3} |u|^2 \diff x = N$, follows from the discussion in \cite{Lieb+Yau1987}; see also \cite{Lenzmann2005LWP} for $m=0$. Furthermore, it turns out that the mass parameter, $m \geq 0$, plays a decisive role summarized as follows; see also \cite{FJL2005} for more details.
\begin{equation*}
\mbox{For $\displaystyle \left \{ {m  > 0 \atop m= 0} \right \}$, $Q(x)$ exists iff $\displaystyle \left \{ { \int_{\RR^3} |Q|^2 \diff x < \Nc  \atop \int_{\RR^3} |Q|^2 \diff x = \Nc } \right \}$, and then $\displaystyle \left \{ {\En(Q)  > 0 \atop \En(Q) = 0} \right \}$.}
\end{equation*} 
Here $\Nc$ is the universal constant appearing in (\ref{eq-Nc}). For $m=0$, we can prove existence of blow-up solutions with initial data $u_0(x)$ arbitrarily close to $Q(x)$; see Theorem \ref{th-instab} below, for a precise statement. In contrast to this instability result, ground state solitary waves turn out to be orbitally {\em stable} whenever $m > 0$ holds; see \cite{FJL2005} for a proof of this fact, as well as a detailed discussion of travelling ground state solitary waves for (\ref{eq-nls-intro}).

Commenting the proof of our main result, we remark that its key ingredient is a virial-type argument for the nonnegative quantity, $M(t)$, defined as
\begin{equation}
M(t) = \int_{\RR^3} \sum_{k=1}^3 \overline{u}(t,x)  x_k \Tm x_k u(t,x) \diff x .
\end{equation}
By restricting to spherically symmetric solutions, $u(t,x)$, of equation (\ref{eq-nls-intro}), various estimates (which become decisively clear when using commutators) and conservation laws, we will derive the inequality
\begin{equation} 
0 \leq M(t) \leq 2 \En(u_0) t^2 + C_1 t + C_2,
\end{equation}
for some finite constants $C_1$ and $C_2$. Thus, if $\En(u_0)$ is negative, we conclude that $u(t,x)$ cannot exist for all times $t \geq 0$, which implies statement (\ref{eq-blowup}), by results taken from \cite{Lenzmann2005LWP}.

It would obviously be of considerable interest to overcome our restriction to spherically symmetric blow-up solutions of (\ref{eq-nls-intro}) and to arrive at a state of affairs comparable to that known for $L^2$-critical nonlinear Schr\"odinger equations; see, \eg, \cite{Cazenave2003} for an overview. Especially, it would be important to get insight into (upper) bounds on blow-up rates and blow-up profiles.

\subsubsection*{Notation}
Throughout this text, $\Lp{p}(\RR^3)$, with norm $\| \cdot \|_p$ and $1 \leq p \leq \infty$, denotes the Lebesgue $L^p$-space of complex-valued functions on $\RR^3$. The complex scalar product on $\Lp{2}(\RR^3)$ is given by
\begin{equation*}
\langle u, v \rangle := \int_{\RR^3} \overline{u} v \diff x .
\end{equation*}
We employ inhomogeneous Sobolev spaces, $\Hs{s}(\RR^3)$, of fractional order $s \in \RR$, which are defined as
\begin{equation*}
\Hs{s}(\RR^3) := \big \{ u \in \mathcal{S}'(\RR^3) : \| u \|_{H^s} := \| \mathcal{F}^{-1} [ 1 + |\cdot|^2 ]^{s/2} \mathcal{F} u \|_2 < \infty \big \},
\end{equation*}
where $\mathcal{F}$ denotes the Fourier transform defined on $\mathcal{S}'(\RR^3)$ (space of tempered distributions). Furthermore, we make use of inhomogeneous Sobolev spaces, $W^{k,\infty}(\RR^3)$, of integer order $k \in \mathbb{N}$, which are given by
\begin{equation*}
W^{k,\infty}(\RR^3) := \big \{ u \in \Lp{\infty}(\RR^3) : \| u \|_{W^{k,\infty}} := \sum_{ \alpha \in \mathbb{N}^3, |\alpha| \leq k} \| \partial^{\alpha} u \|_\infty < \infty \big \}. 
\end{equation*}
The space of complex-valued, smooth functions on $\RR^3$ with compact support is denoted by $\Ctest(\RR^3)$. Moreover, the operator $\sqrt{-\Delta + m^2}$ is defined via its symbol $\sqrt{k^2 + m^2}$ in Fourier space, and the letter $C$ appearing in inequalities denotes constants.


\section{Main Results}
\label{sec-main}

Generalizing equation (\ref{eq-nls-intro}), we introduce the following initial value problem
\begin{equation} \label{eq-ivp}
\left \{ \begin{array}{l} \displaystyle i \dt u = \big ( \Tm  + V \big ) u - \big ( \frac{e^{-\mu |x|}}{|x|} \ast |u|^2 \big ) u, \\
u(0,x) = u_0(x), \quad u : [0,T) \times \RR^3 \rightarrow \CC , \end{array} \right .
\end{equation}
where $m \geq 0$ and $\mu \geq 0$ are parameters, and $V : \RR^3 \rightarrow \RR$ denotes an external potential. We recall from \cite{Lenzmann2005LWP} that, under fairly general assumptions on $V(x)$, we have local well-posedness of $(\ref{eq-ivp})$ in energy space 
\begin{equation} \label{def-X}
X := \big \{ u \in \Hhalf(\RR^3) : V |u|^2 \in \Lp{1}(\RR^3) \big \} .
\end{equation}  
This means that, for any $u_0 \in X$, there exists a unique solution, $u \in C^0( [0,T); X)$, of (\ref{eq-ivp}) with maximal time of existence $T \in (0,\infty]$. In addition, we have conservation of charge (or mass), $\Nn(u)$, and energy, $\En(u)$, which are given by
\begin{equation}
\Nn(u) = \int_{\RR^3} |u|^2 \diff x,
\end{equation}
\begin{align} 
\En(u) & = \phantom{-}\frac{1}{2} \int_{\RR^3} \overline{u} \Tm u \diff x + \frac{1}{2} \int_{\RR^3} V |u|^2 \diff x \nonumber \\
& \quad - \frac{1}{4} \int_{\RR^3} \big ( \frac{e^{-\mu |x|}}{|x|} \ast |u|^2 \big ) |u|^2 \diff x . \label{def-En}
\end{align}

\subsection{Blow-Up}

We now consider external potentials, $V(x)$, that satisfy the following condition.
\begin{assumption} \label{ass-1}
Suppose that $V \in W^{1,\infty}(\RR^3)$ is real-valued with $|V(x)| \leq C ( 1 + |x|)^{-1}$ and $|\nabla V(x)| \leq C (1 + |x|)^{-2}$, for some finite constant $C$.  \end{assumption}
\noindent
In view of (\ref{def-X}), Assumption \ref{ass-1} obviously implies that the energy space, $X$, equals $\Hhalf(\RR^3)$. 

Our first main result shows that spherically symmetric initial data, $u_0(x) \in \Ctest(\RR^3)$, with sufficiently negative energy lead to blow-up of $u(t,x)$ within a finite time.

\begin{theorem} \label{th-blowup}
Let $m \geq 0$, $\mu \geq 0$, and suppose that $V(x)$ is spherically symmetric and satisfies Assumption \ref{ass-1}. Furthermore, we define the bounded function $U := \min \{ V + x \cdot \nabla V, 0 \}$. 

Then any spherically symmetric initial datum, $u_0(x) \in \Ctest(\RR^3)$, satisfying
\begin{equation} \label{ineq-blowup}
 \En(u_0) < - \frac{1}{2} \| U \|_\infty \| u_0 \|_2^2 
\end{equation}
gives rise to a solution, $u(t,x)$, of (\ref{eq-ivp}) that blows up within a finite time, \ie, we have that
\begin{equation}
 \lim_{t \nearrow T} \| u(t, \cdot) \|_{H^{1/2}} = \infty , \quad \mbox{for some $0 < T < \infty$.}
\end{equation}
\end{theorem}

\begin{remarks*}
1) The condition that $u_0(x)$ belongs to $\Ctest(\RR^3)$ can be relaxed to weaker regularity and decay assumptions. But for simplicity of our presentation, we do not pursue this point in more detail.

2) By scaling properties of $\En(u)$, it is easy to see that condition (\ref{ineq-blowup}) always holds for sufficiently large initial data. 

3) In physical terms, Theorem \ref{th-blowup} tells us that the universal constant $\Nc > 4/\pi$ appearing in (\ref{eq-Nc}) can be viewed as a {\em ``Chandrasekhar limit mass''} for a boson star whose dynamics is modelled by equation (\ref{eq-ivp}), where $\mu=0$ and $V \equiv 0$. Indeed, referring to the exposition in \cite{FJL2005} we see that, for every $N > \Nc$, there exists a spherically symmetric initial datum, $u_0(x) \in \Ctest(\RR^3)$, with $\| u_0 \|_2^2 = N$ and such that $\En(u_0) < 0$ holds. By Theorem \ref{th-blowup}, the corresponding solution of (\ref{eq-ivp}), with $\mu=0$ and $V \equiv 0$, blows up within a finite time.     

4) Theorem \ref{th-blowup} can be viewed as a quantum-mechanical extension of the blow-up result derived by \cite{Glassey+Schaeffer1985} for the {\em relativistic Vlasov--Poisson system} which models classical, relativistic particles with Newtonian gravitational interactions.

\end{remarks*}

\subsection{An Instability Result}

We now set $m = \mu = 0$ and $V \equiv 0$ in (\ref{eq-ivp}). As mentioned in Sect.~1, we can address the problem of stability of ground state solitary waves (at rest), 
\begin{equation} \label{eq-sol}
u_{\mathrm{sol}}(t,x) = e^{i\omega t} Q(x) .
\end{equation} 
Recall that the ground state $Q(x)$ has to solve the nonlinear equation 
\begin{equation} \label{eq-Q}
\T \, Q - \big ( \frac{1}{|x|} \ast |Q|^2 \big ) Q = - \omega Q,
\end{equation}
see Sect.~1 and references given there, for existence and spherical symmetry of $Q(x)$. 

We have the following instability result.

\begin{theorem} \label{th-instab}
Set $m=\mu=0$ and $V \equiv 0$ in (\ref{eq-ivp}). Then ground state solitary waves at rest are unstable in the following sense. For any $\eps > 0$, there exists $u_0 \in \Ctest(\RR^3)$ such that $\| u_0 - Q \|_{H^{1/2}} < \eps$ holds and the corresponding solution, $u(t,x)$, of (\ref{eq-ivp}) blows up within a finite time.
\end{theorem}

\begin{remark*}
In contrast to this result, we mention that travelling ground state solitary waves are orbitally {\em stable} whenever the mass parameter is positive, \ie, $m> 0$ holds in (\ref{eq-ivp}). We refer to \cite{FJL2005} for this result.
\end{remark*}

\section{Proof of Main Results}
\label{sec-proofs}

\subsection{Proof of Theorem \ref{th-blowup}}

The proof of Theorem \ref{th-blowup} is organized in four steps as follows.

\subsubsection*{Step 1: Preliminaries} 

We begin with an observation that immediately follows from the field equation (\ref{eq-ivp}): Let $B$ be some time-independent operator on $\Lp{2}(\RR^3)$ and let $u(t)$ be a solution of (\ref{eq-ivp}). Then the time derivative of the expected value, 
\begin{equation}
\inner{ u(t)}{ B u(t)},
\end{equation}
of $B$ is (formally, at least) given by {\em Heisenberg's formula:}
\begin{equation} \label{eq-heisen}
\frac{\dd}{\dd t} \langle u(t), B u(t) \rangle = i \langle u(t), [H,B] u(t) \rangle .
\end{equation} 
Here $[H,B] \equiv HB-BH$ denotes the commutator of $B$ with the time-dependent Hamiltonian $H=H(t)$ given by 
\begin{equation} \label{eq-Ht}
H(t) := \Tm + V + \Vs(t), \quad \mbox{where} \quad \Vs(t) := - \big  ( \frac{e^{-\mu |x|}}{|x|} \ast |u(t)|^2 \big ) .  
\end{equation}
If (\ref{eq-heisen}) is applied for purpose of rigorous arguments, we have to verify that expressions such as $\langle u(t), HB u(t) \rangle$ etc.~are well-defined in the case at hand. 

Next, we note that the proof presented in \cite{Lenzmann2005LWP} for local well-posedness of (\ref{eq-ivp}) with $V \equiv 0$ and initial data in $H^s(\RR^3)$, with $s \geq 1/2$, carries over, with only cosmetic changes, for real-valued $V \in W^{1,\infty}(\RR^3)$ and initial data belonging to $H^s(\RR^3)$, with $2 \geq s \geq 1/2$. 

Therefore we have that, for every $u_0 \in H^2(\RR^3)$, there exists a unique solution
\begin{equation} \label{eq-H2}
u \in C^0 \big([0,T); \Hs{2}(\RR^3) \big ) \cap C^1([0,T); \Hs{1}(\RR^3) \big ),
\end{equation} 
where $T > 0$ denotes the maximal time of existence. Moreover, the following blow-up alternative in energy norm holds: Either $T = \infty$ or $T < \infty$ and $\lim_{t \nearrow T} \| u(t) \|_{H^{1/2}} = \infty$. In addition, we have that $|x|^2 u_0(x) \in \Lp{2}(\RR^3)$ implies that
\begin{equation} \label{eq-xx2}
|x|^2 u(t) \in \Lp{2}(\RR^3), \quad \mbox{for $t \in [0,T)$},
\end{equation}
as shown by Lemma \ref{lem-prop} in Appendix A. In what follows, we will make use of the regularity and decay properties of $u(t)$ stated in (\ref{eq-H2}) and (\ref{eq-xx2}).   

Finally, we mention spherical symmetry of $u(t,x)$ follows from (\ref{eq-ivp}) whenever $u_0(x)$ and $V(x)$ exhibit this property. For conservation of charge, $\Nn(u) = \|  u \|_2^2$, (notice that $V$ is real-valued) and energy, $\En(u)$, we refer to \cite{Lenzmann2005LWP}.

\subsubsection*{Step 2: Dilatation Estimate}

The first crucial step in proving Theorem \ref{th-blowup} is to estimate the time evolution for the expected value of the generator of dilatations,
\begin{equation} \label{def-A}
A := \frac{1}{2} ( x \cdot p + p \cdot x ), 
\end{equation}
where, from now on, we employ the following notation
\begin{equation}
p := - i \nabla.
\end{equation}

\begin{lemma} \label{lem-A}
Let the assumptions on $m$, $\mu$, and $V$ stated in Theorem \ref{th-blowup} (except for spherical symmetry of $V$) be satisfied. Furthermore, suppose that $u_0(x) \in \Ctest(\RR^3)$ holds. Then the map $t \mapsto \langle u(t), A u(t) \rangle$ satisfies the inequality
\begin{equation} \label{ineq-A}
\frac{1}{2} \frac{\dd}{\dd t} \langle u(t), A u(t) \rangle \leq  \En(u_0) + \frac{1}{2} \| U \|_\infty \| u_0 \|_2^2,
\end{equation}
for all $0 \leq t < T$, where $U:=\min \{ V + x \cdot \nabla V, 0 \}$.
\end{lemma}

\begin{proof}[Proof of Lemma \ref{lem-A}]
Since $u(t) \in H^2(\RR^3)$ and $|x|^2 u(t) \in \Lp{2}(\RR^3)$ for all $t\in [0,T)$, as discussed in Step 1 above, it is legitimate to apply (\ref{eq-heisen}); notice that, \eg, $|\langle u(t), (x \cdot p) H u(t) \rangle | \leq \| |x| u(t) \|_2 \| u(t) \|_{H^2}$ is finite.

For brevity, we will write $u$, instead of $u(t,x)$, etc. Using the definition of $H$ in (\ref{eq-Ht}), an elementary calculation with commutators leads to 
\begin{equation}
[H, A] =  \frac{-ip^2}{\sqrt{p^2 + m^2}} + i x \cdot \nabla V + i x \cdot \nabla \Vs ,
\end{equation}
which, by Heisenberg's formula (\ref{eq-heisen}), implies that
\begin{equation} \label{eq-At}
\frac{\dd}{\dd t} \inner{u}{A u} = \innerb{u}{\frac{p^2}{\sqrt{p^2 +m^2}} u} - \innerb{u}{( x \cdot \nabla V) u } - \innerb{u}{(x \cdot \nabla \Vs) u} .
\end{equation}     
Notice that, by our assumptions on $V$, we have that $|\inner{u}{(x \cdot \nabla V) u}| \leq C \inner{u}{|x| u}$ is finite. 

Next, we observe the following identity (using Fubini's theorem and interchanging differentiation and integration, which can be justified easily) 
\begin{align}
\innerb{u}{ (x \cdot \nabla \Vs) u}  & = \int_{\RR^3 \times \RR^3} x \cdot \Big ( \frac{e^{-\mu |x-y|}}{|x-y|^2} \frac{x-y}{|x-y|} \nonumber \\
& \qquad + \mu \frac{e^{-\mu |x-y|}}{|x-y|} \frac{x-y}{|x-y|} \Big )  |u(t,x)|^2 |u(t,y)|^2 \diff x \diff y \nonumber \\
& = \frac{1}{2} \int_{\RR^3 \times \RR^3} \big ( \frac{e^{-\mu |x-y|}}{|x-y|} + \mu e^{-\mu |x-y|} \big )  |u(t,x)|^2 |u(t,y)|^2 \diff x \diff y  \nonumber \\
& = -\frac{1}{2} \inner{u}{ \Vs u} + \frac{\mu}{2} \innerb{ u}{ \big ( e^{-\mu |x|} \ast |u|^2 \big ) u}, 
\end{align}
where the second equation follows from a simple symmetry argument. Furthermore, thanks to the obvious fact that $p^2/(p^2+m^2)^{1/2} = (p^2 + m^2)^{1/2} - m^2/(p^2+m^2)^{1/2}$, we find that (\ref{eq-At}) can be expressed as follows
\begin{align}
\frac{\dd}{\dd t} \inner{u}{A u}  & =  2\En(u_0)  - \innerb{u}{( V + x\cdot \nabla V) u}  \nonumber \\
& \quad -\innerb{u}{\frac{m^2}{\sqrt{p^2+m^2}} u} - \frac{\mu}{2} \innerb{ u}{ \big ( e^{-\mu |x|} \ast |u|^2 \big ) u}  , \label{eq-A2} 
\end{align}
where we use that $\En(u_0)$ is conserved and given by (\ref{def-En}). Since the last two terms in (\ref{eq-A2}) are nonpositive, we can deduce inequality (\ref{ineq-A}) by applying H\"older's inequality and noticing that 
\begin{equation}
-\inner{u}{(V+ x \cdot \nabla V) u} \leq \| U \|_\infty \| u_0 \|_2^2 ,
\end{equation} 
where $U := \min \{ V + x \cdot \nabla V, 0 \}$. This completes our proof of Lemma \ref{lem-A}. \end{proof}

\subsubsection*{Step 3: Variance-Type Estimate}

For our next step in the proof of Theorem \ref{th-blowup}, we introduce 
\begin{equation}
M := x \sqrt{-\Delta+m^2}  x := \sum_{k=1}^3 x_k \sqrt{-\Delta + m^2} x_k ,
\end{equation}
for $m \geq 0$. Note that $M$ is nonnegative, \ie, we have that $\inner{u}{M u} \geq 0$. The time evolution of its expected value is estimated by the following lemma, whose proof rests on the restriction to initial data and external potentials that are spherically symmetric.

\begin{lemma} \label{lem-M}
Let the assumptions on $m$, $\mu$, and $V$ stated in Theorem \ref{th-blowup} be satisfied, and suppose, further, that $u_0(x) \in \Ctest(\RR^3)$ is spherically symmetric. Then the map $t \mapsto \langle u(t), M u(t) \rangle$ satisfies the inequality
\begin{equation} \label{ineq-M}
\frac{1}{2} \frac{\dd}{\dd t} \inner{ u(t)}{ M u(t)} \leq \inner{ u(t)}{ A u(t)} + C,
\end{equation}
for all $0 \leq t < T$, where $C$ is some constant depending only on $\| u_0 \|_2^2$ and $V$.
\end{lemma}

\begin{proof}[Proof of Lemma \ref{lem-M}]
As in the proof of Lemma \ref{lem-A}, we also write $u$ instead of $u(t,x)$, etc.

Since $u(t) \in H^2(\RR^3)$ and $|x|^2 u(t) \in \Lp{2}(\RR^3)$ for all $t \in [0,T)$, by Step 1 discussed above, Heisenberg's formula (\ref{eq-heisen}) can be applied in a rigorous way. To see this, we note, \eg, that $[x, \sqrt{p^2+m^2}] = ip/\sqrt{p^2+m^2}$ holds, which shows that $|\inner{u}{M \sqrt{p^2 + m^2} u} | \leq \| |x|^2 u \|_2 \| u \|_{H^2} + \| |x| u \|_2 \| u \|_{H^1}$ is finite.

Applying now (\ref{eq-heisen}), we find that
\begin{equation} \label{eq-M1}
\frac{\dd}{\dd t} \inner{u}{ M u}  = i \innerb{ u}{ [\sqrt{p^2+m^2}, M] u} + i \innerb{u}{[V,M]u} + i \innerb{u}{[\Vs,M] u} ,
\end{equation}
where $\Vs$ is defined in (\ref{eq-Ht}). Next, we note the identity
\begin{align} 
[V, M] & = [V, x \sqrt{p^2+m^2} x] = Vx \sqrt{p^2+m^2} x - x \sqrt{p^2+m^2} x V \nonumber \\
& = [V x^2, \sqrt{p^2 + m^2} ] - \frac{ip}{\sqrt{p^2+m^2}} \cdot xV - Vx \cdot \frac{ip}{\sqrt{p^2+m^2}} \label{eq-VM},
\end{align}
which is, of course, also true if $V$ is replaced by $\Vs$. Using (\ref{eq-VM}) and applying Lemma \ref{lem-commutator} in Appendix B, we conclude that the first commutator in (\ref{eq-VM}) is a bounded operator on $L^2(\RR^3)$ with norm
\begin{equation}
\big \| \big [V x^2, \sqrt{p^2 + m^2} \big ] \big \|_{L^2 \rightarrow L^2} \leq C \| \nabla ( V x^2 ) \|_\infty \leq C_V,
\end{equation}
for some finite constant $C_V$, thanks to Assumption \ref{ass-1} for $V(x)$. Furthermore, by noticing that $ip/(p^2+m^2)^{1/2}$ is a bounded operator on $L^2(\RR^3)$ and that $\| V x \|_\infty \leq C$ holds (which follows from Assumption \ref{ass-1}), we deduce that the remaining commutators in (\ref{eq-VM}) are also bounded operators on $L^2(\RR^3)$. Thus, we can estimate
\begin{equation} \label{ineq-Vcom}
| \inner{u}{[V,M] u} | \leq C_V \| u \|_2^2 = C_V \| u_0 \|_2^2 ,
\end{equation} 
by conservation of the $L^2$-norm, with some finite constant $C_V$. 

To find an estimate for $[\Vs,M]$ similar to (\ref{ineq-Vcom}), we now show that 
\begin{equation} \label{ineq-V}
\| \nabla (x^2 \Vs) \|_\infty + \| |x| \Vs \|_\infty \leq C_0 \| u_0 \|_2^2,
\end{equation}
with some universal constant $C_0$. In fact, we prove the pointwise estimates
\begin{equation} \label{ineq-V2}
| \Vs(t,x) | \leq \frac{ \| u_0 \|_2^2}{|x|}, \quad | \nabla \Vs(t,x) | \leq \frac{\| u_0 \|_2^2}{|x|^2},
\end{equation}
for all $x \in \RR^3$ and $0 \leq t < T$, which obviously imply (\ref{ineq-V}). 

To prove (\ref{ineq-V2}), we recall from (\ref{eq-Ht}) that $\Vs=\Vs(t,x)$ is given by
\begin{equation} \label{eq-V}
\Vs(t,x) = - \int_{\RR^3} \frac{e^{-\mu |x-y|}}{|x-y|} |u(t,y)|^2 \diff y ,
\end{equation}
Since $u=u(t,x)$ is spherically symmetric (and hence so is $\Vs(t,x)$), we can write $u(t,r)$ and $\Vs(t,r)$ for $r=|x|$ with some abuse of notation. For the moment, let us assume that $\mu=0$ in (\ref{eq-V}). Then, due to the spherical symmetry of $u=u(t,r)$, we can invoke {\em Newton's theorem} (see, \eg, \cite[Theorem 9.7]{Lieb+Loss2001}), which yields the bound
\begin{equation}
|\Vs(t,r)| \leq \frac{1}{r} \int_{\RR^3} |u(t,|y|)|^2 \diff y = \frac{\| u_0 \|_2^2}{r},
\end{equation} 
by conservation of the $L^2$-norm. Moreover, we have the following explicit formula
\begin{equation}
\Vs(t,r) = -\frac{1}{r} \int_{|y| \leq r} |u(t,|y|)|^2 \diff y - \int_{|y| > r} \frac{ |u(t,|y|)|^2}{|y|} \diff y.
\end{equation} 
Taking the derivative with respect to $r=|x|$ (which is allowed by the regularity properties of $|u(t,r)|^2$), we obtain that
\begin{equation}
\frac{\partial}{\partial r} \Vs(t,r) = \frac{1}{r^2}  \int_{|y| \leq r} |u(t,|y|)|^2 \diff y. 
\end{equation}
Hence we conclude that
\begin{equation}
|\nabla \Vs(t,x)| = | \frac{\partial}{\partial r} \Vs(t,r) | \leq \frac{1}{r^2} \int_{|y| \leq r} |u(t,|y|)|^2 \diff y \leq \frac{\| u_0 \|_2^2}{r^2} .
\end{equation}
This proves estimate (\ref{ineq-V2}) for $\mu = 0$.

It remains to show (\ref{ineq-V2}) for $\mu > 0$. Here we apply the following simple trick: We see that $\Vs$ satisfies the Yukawa equation, $(\Delta - \mu^2) \Vs = 4 \pi |u|^2$, which can be rewritten as follows
\begin{equation}
\Delta \Vs = 4 \pi  |u|^2  + \mu^2 \Vs, \quad \mbox{with $\Vs \rightarrow 0$ as $|x| \rightarrow \infty$}.
\end{equation}
By linearity, we have that $\Vs = V_1 + V_2$, where $\Delta V_1 = 4 \pi |u|^2$ and $\Delta V_2 = \mu^2 \Vs$, with $V_1$ and $V_2$ vanishing at infinity. Since $\Vs \leq 0$ by (\ref{eq-V}), we have that $V_2 \geq 0$. Estimating $V_1$ has already been accomplished above. To bound $V_2$ and $\nabla \Vs$, we proceed in the same way replacing only $|u|^2$ by $\frac{\mu^2}{4 \pi} \Vs$. This yields
\begin{equation}
|V_2(t,r)| \leq \frac{1}{r} \frac{\mu^2}{4 \pi} \int_{\RR^3} |\Vs(t,|y|)| \diff y \leq \frac{1}{r} \frac{\mu^2}{4 \pi} \big \| \frac{e^{-\mu r}}{r} \|_1 \| u \|_2^2 = \frac{\| u_0 \|_2^2}{r},  
\end{equation}
where we use Young's inequality and the fact $\int_{\RR^3} e^{-\mu r}{r}^{-1} \diff x = 4 \pi / \mu^2$ whenever $\mu > 0$. Similarly, we find that
\begin{equation}
| \frac{\partial}{\partial r} V_2(t,r) | \leq \frac{\| u_0 \|_2^2}{r^2} . 
\end{equation}
Finally, we note that $V_1 \leq \Vs \leq V_2$ holds, as well as $\frac{\partial}{\partial r} V_2 \leq \frac{\partial}{ \partial r} \Vs \leq \frac{\partial}{\partial r} V_1$, and we eventually arrive at (\ref{ineq-V}). 

Returning to (\ref{eq-M1}), we may now proceed as follows
\begin{equation}
\frac{\dd}{\dd t} \inner{u}{Mu} \leq i \inner{u}{[\sqrt{p^2+m^2},M]u} + C,
\end{equation}
for some constant $C$ depending only on $V$ and on $\| u_0 \|_2^2$. To complete our proof of Lemma \ref{lem-M}, we note that, by a simple calculation, 
\begin{align}
[\sqrt{p^2+m^2}, M] & = [\sqrt{p^2+m^2}, x \sqrt{p^2+m^2} x] \nonumber \\
& = -i( x \cdot p + p \cdot x) = -2i A,
\end{align}
where $A$ is the generator of dilatations introduced in (\ref{def-A}). This completes our proof of Lemma \ref{lem-M}. \end{proof}

\subsubsection*{Step 4: Completing the Proof of Theorem \ref{th-blowup}}

The assertion of Theorem \ref{th-blowup} follows by integrating and combining the differential inequalities (\ref{ineq-A}) and (\ref{ineq-M}) stated in Lemma \ref{lem-A} and \ref{lem-M}, respectively. We find that the nonnegative quantity, $\inner{u(t)}{M u(t)}$, satisfies
\begin{equation}
 \inner{u(t)}{M u(t)} \leq 2 \big ( \En(u_0) + \frac{1}{2} \| U \|_\infty \| u_0 \|_2^2 ) \big )  t^2 + C_1 t + C_2, \quad \mbox{for $t \in [0,T)$},
\end{equation}
where $C_1$ and $C_2$ are some finite constants. If $\En(u_0) < - \frac{1}{2} \| U \|_\infty \| u_0 \|_2^2$ holds, we conclude that the maximal time of existence obeys $T < \infty$. By the blow-up alternative proved in \cite{Lenzmann2005LWP}, this implies that $\| u(t,\cdot) \|_{H^{1/2}} \rightarrow \infty$, as $t \nearrow T$. This completes the proof of Theorem \ref{th-blowup}.

\subsection{Proof of Theorem \ref{th-instab}}

We set $m = \mu = 0$ and $V \equiv 0$ in (\ref{eq-ivp}). As recalled in the introduction (see references given there), we have that any ground state $Q(x)$ is spherically symmetric with respect to some point $x_0$, which we can assume to be the origin. Recalling that $\En(Q) = 0$ holds, we see by elementary arguments that
\begin{align}
\En((1+\delta) Q ) & = \En(Q) + \langle \En'(Q), \delta Q \rangle_{H^{-1/2}, H^{1/2}} + \mathcal{O}(\delta^2) \nonumber \\
& =  - \delta \omega \Nn(Q) + \mathcal{O}(\delta^2) < 0,
\end{align}
for $\delta > 0$ sufficiently small. By density of $\Ctest(\RR^3) \subset \Hhalf(\RR^3)$ and by continuity of $\En$, there exists $u_0 \in \Ctest(\RR^3)$ such that $\En(u_0) < 0$, and $u_0$ is arbitrarily close (in $H^{1/2}$-norm) to $(1+\delta) Q$. Thus, given any $\eps > 0$, we can choose $\delta > 0$ sufficiently small and $u_0 \in \Ctest(\RR^3)$ in such a way that $\| u_0 - Q \|_{H^{1/2}} < \eps$ and $\En(u_0) < 0$ holds. Since $u_0$ is spherically symmetric and $u \in \Ctest(\RR^3)$, we can invoke Theorem \ref{th-blowup} to find that $u(t)$ blows up within finite time. This completes the proof of Theorem \ref{th-instab}.








\begin{appendix}


\section{Propagation of Moments in $x$-Space}

\begin{lemma} \label{lem-prop}
Let $m \geq 0$, $\mu \geq 0$ in (\ref{eq-ivp}), and suppose that $V$ and $u_0(x)$ satisfy the assumptions of Theorem \ref{th-blowup} (except for spherical symmetry). Then  we have that $|x|^2 u(t) \in \Lp{2}(\RR^3)$ holds on $[0,T)$.
\end{lemma} 

\begin{proof}[Proof of Lemma \ref{lem-prop}]
First, we show that the second moment of $u(t)$ in $x$-space stays finite along the flow, \ie,  
\begin{equation} \label{eq-x2}
|x| u(t) \in \Lp{2}(\RR^3), \quad \mbox{on $[0,T)$.}
\end{equation} 
To prove this claim, we introduce the regularized quantity
\begin{equation} \label{eq-f}
f_\eps(t) := \innerb{u(t)}{|x|^2 e^{-2\eps|x|} u(t)} ,
\end{equation}
where $\eps > 0$. For $f_\eps(t)$ and $t \in [0,T)$, a well-defined calculation yields   
\begin{align}
f_\eps(t) & = f_\eps(0) + \int_0^t f_\eps'(s) \diff s  \nonumber \\ 
& = f_\eps(0) + i \int_0^t \big \langle u(s), [\Tpm, |x|^2 e^{-2\eps |x|} ] u(s) \big \rangle \diff s \nonumber \\
& = f_\eps(0) + i \int_0^t \big \langle u(s), x e^{-\eps |x|} \cdot [\Tpm, x e^{-\eps |x|} ] \nonumber \\
& \quad + [\Tpm, x e^{-\eps |x|} ] \cdot x e^{-\eps |x|} u(s) \big \rangle \diff s . \label{eq-f2}
\end{align}   
Invoking the Cauchy-Schwarz inequality and noting that $\sqrt{f_\eps(t)} = \| |x| e^{-\eps |x|} u(t)\|_2$, we obtain 
\begin{equation} \label{ineq-fe}
f_\eps(t) \leq f_\eps(0) + C \int_0^t \big \| \big [\Tpm, |x| e^{-\eps |x|} \big ] u(s) \big \|_2 \sqrt{f_\eps(s)} \diff s .
\end{equation}
By Lemma \ref{lem-commutator} in Appendix A.2 and the fact that $|x| e^{-\eps |x|}$ has a uniformly bounded gradient, we have that
\begin{equation} 
\big \| \big [\Tpm, |x| e^{-\eps |x|} \big ] \big \|_{L^2 \rightarrow L^2} \leq C \big \| \nabla \big ( |x| e^{-\eps |x|}  \big ) \big \|_\infty \leq C.
\end{equation}
Thus, it is a bounded operator from $\Lp{2}(\RR^3)$ into itself (and the bound on its operator norm is independent of $\eps$). 
 
Let us now return to (\ref{ineq-fe}). Appealing to conservation of the $L^2$-norm of $u(t)$, we conclude that
\begin{equation} 
f_\eps(t) \leq f_\eps(0) + 2C \int_0^t \sqrt{f_\eps(s)} \diff s,
\end{equation}
This estimate is easily seen to imply that
\begin{equation}
\sqrt{f_\eps(t)} \leq f_\eps(0) + C \int_0^t \diff s = f_\eps(0) + Ct .
\end{equation}
By monotone convergence, we can take the limit $\eps \rightarrow 0$, which leads to
\begin{equation} \label{ineq-f2}
\sqrt{f(t)} \leq f(0) + Ct ,
\end{equation}
using that $f(0) = \| |x| u_0 \|_2 \leq C( \| u_0 \|_2 + \| |x|^2 u_0 \|_2 )$ is finite because $u_0$ and $|x|^2 u_0$ belong to $\Lp{2}(\RR^3)$. This proves (\ref{eq-x2}).

Next, we prove a similar statement for the fourth moment, \ie, 
\begin{equation}
|x|^2 u(t) \in \Lp{2}(\RR^3), \quad \mbox{on $[0,T)$.}
\end{equation}
This follows in fact from (\ref{eq-x2}), as we now show. In analogy to (\ref{eq-f}), we define 
\begin{equation}
g_\eps(t) := \innerb{u(t)}{|x|^4 e^{-2 \eps|x|} u(t)},
\end{equation}
where $\eps > 0$. A calculation similar to (\ref{eq-f2}) yields that
\begin{equation}  \label{ineq-g}
g_\eps(t) \leq g_\eps(0) + 2C \int_0^t \big \| \big [ \Tpm, |x|^2 e^{-\eps |x|} \big ] u(s) \big  \|_2 \sqrt{g_\eps(s)} \diff s .
\end{equation}
Further evaluation of the commutator leads to
\begin{align}
\big \| \big [ \Tpm, |x|^2 e^{-\eps |x|}  \big ] u \big \|_2 & = \big \| \big ( e^{-\eps |x|} x \cdot [\Tpm, x] \nonumber \\
& \qquad  + \big [ \Tpm, x e^{-\eps |x|} \big ] \cdot x  \big ) u \big \|_2 \nonumber \\
& \leq C \big ( \big \| x \cdot \frac{p}{\sqrt{p^2+m^2}} u \big \|_2 + \| |x| u \|_2 \big ) , \label{ineq-com2}
\end{align} 
using that $[\sqrt{p^2+m^2},x]$ and $[\sqrt{p^2+m^2}, x e^{-\eps|x|}]$ are bounded operators on $\Lp{2}(\RR^3)$, see Lemma \ref{lem-commutator}. Next, by a simple commutator calculation, we find that
\begin{equation}
x \cdot \frac{p}{\sqrt{p^2+m^2}} = \frac{p}{\sqrt{p^2+m^2}} \cdot x + i \left ( \frac{3}{(p^2+m^2)^{1/2}} - \frac{p^2}{(p^2+m^2)^{3/2}} \right ) .
\end{equation}
Hence the first term on the right-hand side of (\ref{ineq-com2}) can be estimated as follows
\begin{equation} \label{ineq-x2}
 \| x \cdot \frac{p}{\sqrt{p^2+m^2}} u \big \|_2  \leq C \big ( \big \| \frac{p}{\sqrt{p^2 + m^2}} \cdot x u \big \|_2 + \big \| \frac{1}{|p|} u \big \|_2 \big ) \leq C \| |x| u \|_2, 
\end{equation}
using Hardy's inequality, $\| |x|^{-1} u \|_2 \leq C \| |p| u \|_2$, which also holds when $p$ and $x$ are interchanged, by Fourier duality.

Commenting the proof given above for (\ref{ineq-x2}), we mention that we could also control appeal to weighted $L^2$-estimates for the singular integral operator, $T = p/\sqrt{p^2+m^2}$, and noting that the weight $\omega(x) = |x|^2$ belongs the class $A_2$; see, \eg, \cite[Section V.5]{Stein1993}.  

In summary, we conclude that estimates (\ref{ineq-g}) and  (\ref{ineq-x2}) imply that
\begin{equation}
g_\eps(t) \leq g_\eps(0) + 2C \int_0^t \big ( \|u(s)\|_2 + \||x| u(s)\|_2 \big ) \sqrt{g_\eps(s)} \diff s .
\end{equation}
By conservation of $\|u(t)\|_2$ and estimate (\ref{ineq-f2}) for $\sqrt{f(t)} = \| |x| u(t) \|_2$, we deduce, in a similar fashion as above, that 
\begin{equation}
\sqrt{g_\eps(t)} \leq g_\eps(0) + C \int_0^t ( 1 + s ) \diff s 
\end{equation}
holds. This bound finally leads to
\begin{equation}
\sqrt{g(t)} \leq g(0) + C \big ( t + \frac{t^2}{2} \big ) , 
\end{equation}
when passing to the limit $\eps \rightarrow 0$, by monotone convergence. The proof of Lemma \ref{lem-prop} is now complete. \end{proof}

\section{Commutator Estimate}

\begin{lemma} \label{lem-commutator}
Let $m \geq 0$, and suppose that $f(x)$ is a locally integrable. If the distributional gradient, $\nabla f(x)$, is an $L^\infty(\RR^3)$ vector-valued function, then we have that
\[
 \big \| \big [\Tm, f \big ] \big  \|_{L^2 \rightarrow L^2} \leq C \| \nabla f \|_\infty,
\]
for some constant $C$ independent of $m$.
\end{lemma}

\begin{remark*} Although this result can be deduced by means of Calder\'on--Zygmund theory for singular integral operators and its consequences for pseudo-differential operators (see, \eg, \cite[Section VII.3]{Stein1993}), we present an elementary proof which makes good use of the spectral theorem, enabling us to write the commutator in a convenient way. 
\end{remark*}

\begin{proof}
For $m \geq 0$, we set 
\begin{equation}
A:= \sqrt{p^2 +m^2}, \quad \mbox{where $p := -i \nabla$}.
\end{equation}
Since $A$ is a self-adjoint operator on $\Lp{2}(\RR^3)$ (with domain $\Hs{1}(\RR^3)$), functional calculus (for measurable functions) yields the formula
\begin{equation}
A^{-1} = \frac{1}{\pi} \int_0^\infty \frac{1}{\sqrt{s}} \frac{\diff s}{A^2+s} .
\end{equation}
Due to this fact and $A=A^{-1} A^{2}$, we obtain the formula
\begin{equation}
[ A, f ] = \frac{1}{\pi} \int_0^\infty \frac{\sqrt{s}}{A^2 + s} [A^2, f] \frac{\diff s}{A^2 + s} .
\end{equation} 
Clearly, we have that $[A^2, f] = [p^2,f] = p \cdot [p,f] + [p,f] \cdot p$, which leads to
\begin{equation}
[A,f] = \frac{1}{\pi} \int_0^\infty \frac{\sqrt{s}}{p^2 + m^2 + s} \big ( p \cdot [p,f] + [p,f] \cdot p \big ) \frac{\diff s}{p^2 + m^2 + s} .
\end{equation}
Moreover, since $[p,f] = -i \nabla f$ holds, we have that
\begin{equation}
\big \| \big [ \frac{1}{p^2+m^2+s}, [p,f] \big ] \big \|_{L^{2} \rightarrow L^2}  \leq \frac{2}{s} \| \nabla f\|_\infty .
\end{equation}
Hence we can estimate, for arbitrary test functions $\xi, \eta \in \Ctest(\RR^3)$, as follows:
\begin{align}
& \Big | \innerb{\xi} {\int_0^\infty \frac{\sqrt{s}}{p^2 + m^2 + s} \big ( [p,f] \cdot p \big ) \frac{\diff s}{p^2 + m^2 + s} \, \eta } \Big | \nonumber \\
 & \leq \Big | \innerb{[p,f] \xi}{p \int_0^\infty \frac{\sqrt{s} \diff s}{(p^2+m^2+s)^2} \, \eta} \Big | \nonumber \\
& \quad + \Big | \innerb{\xi}{ \int_0^\infty \big [ \frac{1}{p^2+m^2+s}, [p,f] \big ] \cdot \frac{  p \sqrt{s} \diff s}{p^2+m^2+s} \, \eta } \Big | \nonumber \\
 & \leq \big \| [p,f] \xi \big \|_2 \big \|  \int_0^\infty \frac{p \sqrt{s}  \diff s}{(p^2+m^2+s)^2}  \, \eta \big \|_2  \nonumber \\
& \quad + 2 \| \xi \|_2 \| \nabla f \|_\infty \big \|  \int_0^\infty \frac{ p  \diff s}{\sqrt{s} (p^2+m^2+s)} \, \eta \big \|_2. \label{ineq-L2bound}
\end{align}
Evaluation of the $s$-integrals yields
\begin{align*}
\mbox{r.h.s. of (\ref{ineq-L2bound})} & \leq C \big \| \nabla f\|_\infty \| \xi \big \|_2  \big \| \frac{  p }{\sqrt{p^2+m^2}} \eta \big \|_2   \leq C \| \nabla f \|_\infty  \| \xi \|_2 \| \eta \|_2 .
\end{align*}
The same estimate holds if $[p,f] \cdot p$ is replaced by $p \cdot [p,f]$ in (\ref{ineq-L2bound}). Thus, we have found that
\begin{equation}
\big | \inner{\xi}{ [A,f] \eta}  \big | \leq C \| \nabla f \|_\infty \| \xi \|_2 \| \eta \|_2, \quad \mbox{for $\xi, \eta \in \Ctest(\RR^3)$},
\end{equation}
with some constant $C$ independent of $m$. Since $\Ctest(\RR^3)$ is dense in $L^2(\RR^3)$, the assertion for the $L^2$-boundedness of $[A,f]$ now follows. This completes the proof of Lemma \ref{lem-commutator}.\end{proof}

\end{appendix}






\subsubsection*{Acknowledgments}
We are grateful to D.~Christodoulou, L.~Jonsson, I.~Rodnianski, M.~I.~Sigal, and M.~Struwe for various, valuable discussions. We also thank the referee for some useful comments.




\frenchspacing

\bibliographystyle{plain}


\bigskip
\noindent 
{\sc J\"urg Fr\"ohlich\\
Institute for Theoretical Physics\\
ETH Z\"urich--H\"onggerberg\\
CH-8093 Z\"urich, Switzerland}\\
{\em E-mail address:} {\tt juerg@itp.phys.ethz.ch}
 
\bigskip
\noindent
{\sc Enno Lenzmann\\
Department of Mathematics\\ 
ETH Z\"urich--Zentrum, HG G 33.1\\
R\"amistrasse 101\\
CH-8092 Z\"urich, Switzerland}\\
{\em E-mail address:} {\tt lenzmann@math.ethz.ch}

\end{document}